# Grain boundary-limited thermal transport in suspended thin graphite across an unexplored thickness regime

Wonjae Jeon[1,#], Yu Pei[1,#], Xun Li[2,#], Lucas Lindsay[2], Sangyeop Lee[3,*], Renkun Chen[1,**]

[1]Department of Mechanical and Aerospace Engineering, University of California San Diego, La Jolla, California 92093 USA

[2]Materials Science and Technology Division, Oak Ridge National Laboratory, Oak Ridge, Tennessee, 37831, USA

[3]Department of Mechanical Engineering and Materials Science, University of Pittsburgh, Pittsburgh, Pennsylvania 15261, USA

[#]These authors contributed equally

[*]Correspondence: sylee@pitt.edu

[**]Correspondence: rkchen@ucsd.edu

## TOC Graphic

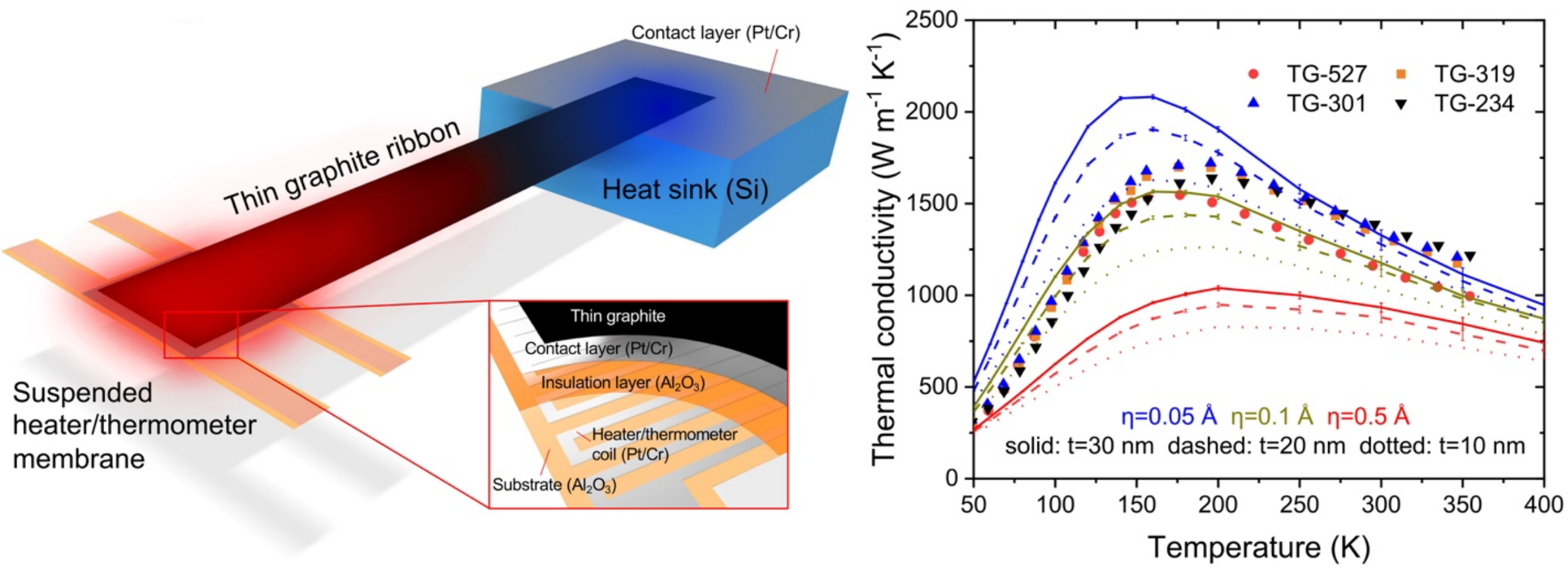

## ABSTRACT

We present systematic thermal conductivity ($\kappa$) measurements of suspended thin graphite ribbons, 234 - 527 nm thick, using a four-probe 3ω method. Unlike recent reports of phonon hydrodynamics and exceptionally high $\kappa$ in micrometer-thick graphite (Science, 2020), we observe significantly lower $\kappa$ and no signatures of collective phonon flow in this intermediate thickness regime. Instead, our measured $\kappa$ lies between few-layer graphene and bulk graphite. These results agree with a first-principles-informed Peierls–Boltzmann transport model with spatially resolved Monte Carlo sampling. Additionally, the temperature for the peak $\kappa$ shifts lower with increasing thickness, due to the interplay of phonon-boundary and phonon-isotope scattering. Incorporating grain boundary scattering into simulations is necessary to replicate the experimental trends. These findings delineate the boundary between ballistic, hydrodynamic, and diffusive transport regimes in graphite, and underscore the dominant role of disorder and geometry in phonon transport in quasi-two-dimensional materials, offering insights for nanoscale thermal management.

Advancements in high-power portable electronic devices and their miniaturization have led to an increased demand for enhanced thermal management to cope with the rising heat fluxes [1, 2]. To meet this demand, materials having high thermal conductivity ($\kappa$) and lightweight are increasingly demanded. Graphene has received considerable attention due to its high in-plane $\kappa$ at room temperature, which is attributed to its strong covalent bonding, light atomic masses[3-6], and peculiarities related to its flat two-dimensional symmetry[7, 8] . It has offered opportunities for novel heat dissipation applications in nanoelectronics. For example, few-layer graphene has been utilized as a heat spreader in high-power gallium nitride (GaN) transistors.[9] However, the small cross-sectional area of single- and few- layer graphene has a limited heat dissipation rate, therefore, a larger number of graphene layers is required to be utilized in the cooling of high-power microdevices[10]. Bulk highly ordered pyrolytic graphite (HOPG), though having a lower in-plane $\kappa$ (1950 W m$^{-1}$ K$^{-1}$)[11] than graphene, has much higher room temperature values than metals such as aluminum (~237 W m$^{-1}$ K$^{-1}$) and copper (~401 W m$^{-1}$ K$^{-1}$)[12] which typically serve as macroscopic heat sinks. As a result, graphite has been extensively studied in various heat dissipation systems, including heat sinks, thermal interface materials, and heat spreaders[13-16].

As heat sinks and spreaders in thermal management systems are fabricated in a wide range of sizes and geometries, it is essential to investigate the size-dependent thermal properties of finite structures [17]. For example, studies on silicon nanowires have shown that thermal conductivity decreases with decreasing diameter due to enhanced phonon-boundary scattering, highlighting the broad spectrum of phonon mean free paths [18]. In $NbSe_3$ nanowires, a non-monotonic dependence of thermal conductivity on diameter has been observed, indicating a competition between phonon-boundary scattering and low-dimensional effects as the diameter is reduced [19].

Recent reports have suggested the emergence of phonon hydrodynamics in thin graphite (TG)

ribbons with micrometer-scale thickness. Notably, an ultrahigh thermal conductivity (κ) of 4300 $W \cdot m^{-1} \cdot K^{-1}$ was measured just below room temperature in a TG ribbon with a width of 350 μm and thickness of 8.5 μm using a thermocouple-based technique [20]. This behavior was attributed to reduced Umklapp (U) scattering, which limits momentum loss during phonon-phonon interactions, thereby enhancing hydrodynamic phonon transport as the sample thickness decreases [20].

However, this interpretation contrasts with findings by Li et al. [21], who conducted Monte Carlo (MC) simulations and experimental measurements on TG samples with thicknesses ranging from 4.4 to 7.7 nm. Their results showed a monotonic decrease in κ with decreasing thickness, consistent with boundary scattering effects, and no signatures of hydrodynamic behavior. The absence of phonon hydrodynamics in their measurements was attributed to the small sample dimensions (width ~1.75 μm) being much smaller than the phonon mean free path for normal (N) scattering along both in-plane and cross-plane directions.

More recently, Nomura and colleagues examined phonon hydrodynamics in graphite samples with 85 nm thickness and found that only isotopically purified samples exhibited hydrodynamic signatures, and only at temperatures below 90 K [22, 23]. These contrasting results underscore the complex interplay between phonon scattering mechanisms, sample purity, and geometry, motivating the present study to re-examine phonon hydrodynamic transport in natural thin graphite.

While the cross-plane thermal conductivity of graphite has been measured across a range of thicknesses—24 to 714 nm by Fu et al. [24], and 90 nm to 21.9 μm by Zhang et al. [25]—revealing long phonon mean free paths along the c-axis, graphite's strong anisotropy necessitates direct measurements of in-plane thermal conductivity over a similarly broad thickness range. Such measurements are critical for understanding phonon transport within the basal planes and for validating models that account for the directional dependence of scattering mechanisms.

We present thermal conductivity (κ) measurements of thin graphite (TG) samples with thicknesses of a few hundred nanometers—a regime not previously explored. Mechanically exfoliated graphite flakes having 352 – 527 nm thicknesses were etched into ribbons of >0.5 mm lengths and ~110 μm widths. They were then transferred onto suspended microdevices designed for a four-probe $3\omega$ $\kappa$ measurement technique. The suspended lengths and width of TG ribbons are 0.5 mm and 0.1 mm respectively. The measured $\kappa$ of TG from 50 to 360 K lies between those of 2.45 nm thickness TG and bulk HOPG.[11, 21] For TG ribbons with thicknesses ranging from 234 nm to 319 nm, $\kappa$ first increases and then decreases with temperature, peaking at around ~1700 W $m^{-1}$ $K^{-1}$ at ~200 K. The temperature dependence of $\kappa$, defined as $\alpha = d(ln\,\kappa\,)/d(ln\ T\,)$, used to examine the existence of phonon hydrodynamics, as it has been reported that $\alpha$ exceeding that value of the ballistic limit is strong evidence of phonon hydrodynamics.[20-22, 26-29].

To further understand the observed trends, we performed first-principles-based Monte Carlo simulations solving the Peierls-Boltzmann equation, incorporating phonon-phonon, boundary, and isotope scattering. The simulations indicate that additional scattering mechanisms, such as grain boundary effects, play a crucial role in determining $\kappa$, particularly for thicker TG samples. Additionally, we identified a thickness-dependent peak temperature for $\kappa$, consistent with prior studies on HOPG. These results provide new insight into phonon transport in quasi-2D graphite and highlight the importance of mesoscopic effects in thermal conduction at intermediate thicknesses, with implications for heat management in microelectronic applications.

To investigate thermal transport in TGs with thicknesses in the hundreds of nanometers range, we exfoliated natural-abundant HOPG (~1.1% $^{13}C$), using an improved tape method to obtain residue-free TG flakes (Supplementary Note S1 and Figure S1). The flakes were then patterned into ribbons using oxygen plasma etching with a shadow mask. Next, a TG ribbon was transferred onto a microfabricated device, as shown in the schematic drawing and scanning electron microscopy (SEM) image in Figure 1a, b. A similar device was used recently for high-conductance microscale samples [30]. Details of the device fabrication process and TG transfer are provided in Supplementary Note S2 and Figures S2 & 3. The suspended length of the TG ribbon ($L_{TG}$) was determined by SEM. The ribbon thickness was determined by scanning the ribbon surface on the heat sink with a surface profilometer (Figure S4), confirms good uniformity. SEM images and surface profiles of all the TG ribbons measured in this study are presented in Figures S5–S8. The dimensions of the TG samples are listed in Table 1.

Figure 1c shows the circuit diagram of the four-probe 3ω method on the suspended devices, with further details of the κ measurement in Supplementary Note S3. We confirm that heat losses due to residual air at the chamber pressure (~$4.5\times10^{-5}$ torr), as well as radiation heat loss, are negligible for the Pt beams and the TG samples (Supplementary Note S4). Using the measured temperature rise of the heater/thermometer membrane ($\Delta T_h$) and the generated heat flux from the heater ($Q_t$), the combined thermal conductance of the TG ($G_{TG}$) and beams ($G_b$) can be calculated (Figure S9)[31-33]. The $\kappa$ of TG was calculated as $\kappa = \frac{G_{TG} \cdot L_{TG}}{A_{TG}}$, where $A_{TG}$ is the cross-sectional area of the TG ribbon. To verify the reliability of our method, we measured the $\kappa$ of a Pt ribbon using the same platform (Supplementary Note S3 and Figure S10).

Figure 2a shows the temperature dependence of $\kappa$ for TGs within the temperature range of 50 and 360 K. The error bars, which are less than ~4%, represent the combined uncertainties from multiple measured data points for $G_{TG}$ and the characterization of TG ribbon dimensions. (Supplementary Note S5).[31-33]. We note that the $\kappa$ measurements presented here correspond to covalently bonded in-plane transport in the TG samples. The peak $\kappa$ value varied in the temperature range from 180 to 205 K for all TG samples. Below the peak temperatures, $\kappa$ decreases with decreasing temperature, characteristic of strong phonon boundary scattering relative to intrinsic phonon-phonon scattering. Above the peak temperatures, phonon-phonon interactions strengthen with increasing temperature, particularly for momentum-destroying U processes, and the thermal conductivity decreases [24, 25]. For TG ribbons with thicknesses ranging from 234 nm to 319 nm, $\kappa$ is around ~1700 W $m^{-1}$ $K^{-1}$ at 200 K (Figure 2a and b). However, these values decrease as the thickness further increases to 527 nm.

The lower $\kappa$ in the thickest sample (TG-527) is quite intriguing. We first examine if it is due to defects. We measured Raman spectra to probe defects in the TGs, as shown in Figure 2c. The presence of a D-band (~1300 $cm^{-1}$) peak indicates the presence of defects in each sample, and its intensity reflects the defect density. Small D-band peaks were observed in TGs with thicknesses ranging from 234 nm to 527 nm, whereas no such peak was detected in bulk HOPG. This suggests that defects were indeed introduced during sample preparation, including graphite exfoliation, ribbon patterning, and TG transfer. Figure 2d shows the ratio of the peak intensity of the D-band to that of the G-band ($I_D/I_G$), based on which the vacancy defect fraction can be estimated to between $1.18 \times 10^{-5}$ and $1.65 \times 10^{-5}$ using a well-established correlation [34, 35]. Even though TG-527 exhibits the highest $I_D/I_G$, as we shall see later, this level of defect fraction does not materially impact the thermal conductivity. A plausible cause for the lower $\kappa$ of TG-527 may be due to the reduced van der Waals interaction between the thicker TG and the suspended device, resulting in increased thermal contact resistance.

We performed first-principles-based theoretical calculations to understand the measured $\kappa$ behaviors of the TG samples. To realistically model the effects of finite sample size on thermal transport, we solved the Peierls-Boltzmann Equation (PBE) in both real and reciprocal spaces. The steady-state PBE with a full scattering matrix is

$$v_i \cdot \nabla_r f_i(r) = \sum_j C_{ij} f_j(r) \quad (1)$$

where $i$ indicates a phonon state, $f$ is the phonon distribution function, $v$ is the phonon group velocity, $r$ is the position vector in three-dimensional real space, and $C_{ij}$ is an element of the scattering matrix $C$. The PBE was solved by a deviational Monte Carlo (MC) method with phonon dispersion and scattering matrix from first-principles calculations as inputs (see Supplementary Note S6 for details). This has been used previously to study thermal transport in two-dimensional graphene ribbons[26, 36-38] and bulk graphite[21, 39], and transient simulation of second sound in three-dimensional graphite[40]. The details of the algorithm can be found in these prior reports. In this study, the discretization of the reciprocal space, real space, and time domain are the same as in Ref. [39].

In Figure 3a, we demonstrate that for infinitely wide ($W$) and thick ($t$) graphite samples, $\kappa$ converges to bulk (infinitely long) values when the sample length is larger than a few hundred micrometers. Considering that the lengths of the samples in the measurements are about 0.5 millimeters, we impose that the deviations of phonon distribution functions from local equilibria are invariant along the heat flow direction, representing a sample with infinite length along this direction. For side surfaces with finite $W$ and $t$, fully diffuse boundary scattering was applied as shown in Figure 3b.

We first calculate the $\kappa$ of thin graphite samples with the size dimensions corresponding to experimental samples TG 234-527 shown in Figure 3c. We consider both pure samples without phonon-isotope scattering and with this scattering arising from natural isotopic mass variations. All calculations overestimate measurements over the full temperature range. This indicates that there are additional scatterings besides phonon-phonon, phonon-isotope, and lateral boundary scatterings in the measurements that provide further thermal resistance. We also examined the effect of point defects and found that point defect scattering alone cannot explain the measured behavior of the thermal conductivity, as detailed in Supplementary Note S7. To examine the existence of phonon Poiseuille flow, we calculate the temperature dependence of $\kappa$, defined as $\alpha = d(ln\,\kappa)/d(ln\ T)$, see Figure 3d. The black dash-dotted line in Fig. 3d is the ballistic limit of phonon transport calculated directly from phonon dispersions representing the temperature dependence of phonon conductance. It has been reported that $\alpha$ exceeding that given by the ballistic limit is strong evidence of phonon hydrodynamics.[20-22, 26-29]. This is because when momentum-conserving normal scattering is strong as in the hydrodynamic regime, phonons experience a random walk trajectory with an effective traveling distance proportional to $W^2/\Lambda_{\mathrm{N}}$ where $W$ is the characteristic sample size and $\Lambda_{\mathrm{N}}$ is the phonon mean free path due to normal scattering. This is contrary to the ballistic case where the mean free path is the characteristic sample size and thus temperature independent. As $\Lambda_{\mathrm{N}}$ decreases with increasing temperature, the effective mean free path increases with temperature, adding a positive temperature component compared to the ballistic case. The calculated $\alpha$ values for all samples are shown in Figure 3d. For TG-527, there is a small temperature window between 60 and 90 K where phonon Poiseuille flow may occur when phonon-isotope scattering is not present. All other samples are not expected to exhibit phonon Poiseuille flow, even if they are isotopically enriched, due to the sample size.[22, 27, 28] Other scattering mechanisms, likely present in our samples, could further reduce the temperature dependence and limit the possibility for hydrodynamic effects.

Electron Back Scattering Diffraction (EBSD) measurements show that in-plane grain sizes are around 70 $\mu m$ (Figure S11), while other studies have shown that out-of-plane grain sizes in HOPG typically vary from 5 to 30 nm[41]. Thus, we investigate the effect of grain boundaries as an additional scattering mechanism when comparing calculations with measurements here. To account for grain boundary resistance, we modify the previous MC simulations by providing a specularity condition to all six surfaces of each sample so that phonon transmission across a boundary is specular or diffuse, depending on a specularity parameter $p$, as shown in the schematic in Figure S12. Each phonon state remains unchanged in specular transmission and is randomly redistributed through diffuse transmission. Periodic boundary conditions are applied so that phonons that transmit across a surface reenter the sample through the opposite surface.

The specularity for phonon state $i$ is calculated as[42] [43]

$$p_i = exp\left(-16\frac{\pi^2}{\lambda_i^2}\eta^2\cos^2\theta\right) \quad (2)$$

where $\lambda$ is the phonon wavelength, $\eta$ is the surface roughness, and $\theta$ is the incident angle. We plot mode specularities for phonons in the first Brillouin zone in Figure S13 for surfaces along different directions and with different roughness. A roughness of 1 Å gives diffuse scattering for most of the phonon states, while a roughness of 0.05 Å leads to partially specular, partially diffuse scattering.

We simulate phonon transport across in-plane grain boundaries in samples with $L = W = 70\ \mu m$, and with varying thickness $t$ and roughness $\eta$. The calculated thermal conductivity $\kappa = \kappa(T, t, \eta)$ is shown in Figure 4a and compared with measurements of TG 234-527. The $\theta$ increases with increasing thickness and decreasing roughness. Assuming a roughness between 0.05 – 0.1 Å for the out-of-plane grain boundaries, the out-of-plane grain size of ~ 30 nm best describes the measurements for all samples.

The $\kappa$ values of TGs with thicknesses ranging from 234 to 527 nm were compared with those from other HOPGs with thicknesses ranging from 2.45 nm to 1 mm (bulk) (Figure 4b).[11, 20-22, 44] The measured $\kappa$ values of TGs with 234 to 527 nm thickness are larger than those of 2.45 and 7.7 nm thick exfoliated HOPG, where the momentum-destroying boundary scattering dominates due to the thickness being smaller than the c-axis phonon MFPs,[24, 25] while it was smaller than that of bulk HOPG with a thickness of 1 mm, indicating a thickness-dependence of $\kappa$. This trend is consistent with MC simulations, which show a higher $\kappa$ as the thickness increases from several nanometers to 65 μm.[21] In contrast, the $\kappa$ measured by the thermocouple method increases as the thickness decreases from 580 to 8.5 μm, exhibiting the opposite trend.[20] At temperatures above 150 K, the $\kappa$ of 8.5 μm thick graphite is even higher than that of bulk HOPG, reaching up to 4300 W $m^{-1}$ $K^{-1}$ at room temperature. This can be attributed to the large N-dominated scattering phonon MFPs, which are comparable to the 8.5 μm thickness at low temperature (50 K), where phonon Poiseuille flow can occur.[21] However, at room temperature, the N-scattering-dominated MFPs are significantly smaller than the micrometer scale and phonon hydrodynamics is barely present.[21] Moreover, the presence of the isotopes in the natural graphite (~1.1% $^{13}C$ in HOPG) can hinder phonon hydrodynamics due to resistive phonon-isotope scattering, which is a momentum-destroying process.[22, 27, 28] A comparative study by Nomura and coworkers on isotopically purified (0.02% $^{13}C$) and natural graphite (1.1% $^{13}C$) reported that natural graphite with a thickness of 85 nm exhibits lower $\kappa$ than the TGs studied in this work (234 - 527 nm thick) (Figure 4b)[22], presumably due to the stronger boundary scattering in the thinner sample. In contrast, isotopically purified graphite with the same thickness (85 nm) shows higher $\kappa$ than the TGs studied here, despite its smaller thickness, demonstrating the suppressive role of isotope-induced phonon scattering on phonon hydrodynamics. Therefore, further studies in a similar thickness scale and the isotope effect of TG are needed to interpret the ultra-high $\kappa$ of 8.5 μm thick HOPG, which surprisingly exceeds that of bulk HOPG.

Figure 4c shows the $\kappa$ versus $T$ peak temperatures for HOPG samples of varying thicknesses.[11, 20, 21, 44] The results of this work show that the peak temperature decreases from 205.3 K to 181.4 K as the thickness increases from 234 nm to 527 nm. These peak temperatures are lower than those of 2.45 nm (292 K) and 7.7 nm (286 K) thick HOPG samples, while these are higher than that of the bulk HOPG (1 mm thick) (118 K), demonstrating a clear dependence of the peak temperature on thickness.[11, 21, 44] The peak temperatures for HOPG samples with thicknesses of 8.5, 240, and 580 μm, measured using the thermocouple method, also decrease with increasing thickness, although their peak temperature range overlaps with the temperatures studied in this work. Further experimental investigation in a thickness range from a few hundred nanometers to a few hundred micrometers is required to fully understand the thickness-dependence of the peak temperature.[21]

In summary, we systematically investigated thermal transport in suspended TG ribbons with thicknesses ranging from 234 to 527 nm using a suspended four-probe 3ω method. The samples were mechanically exfoliated and integrated onto microfabricated devices for accurate in-plane thermal transport measurements. The measured $\kappa$ from 50 to 360 K lies between those reported for few-layer graphene and HOPG) showing clear thickness dependence. The results reveal that $\kappa$ increases with thickness, consistent with spatially resolved Monte Carlo solutions to the Peierls-Boltzmann transport equation. Furthermore, the temperature at which $\kappa$ versus temperature peaks decreases as the graphite thickness increases, indicating evolving phonon scattering behavior. Notably, phonon Poiseuille flow was not observed in this thickness range for our samples with natural isotope abundance. This is likely due to dominant momentum-destroying scattering processes, such as boundary scattering and phonon-isotope scattering from the natural $^{13}C$ content (~1.1%) in HOPG.[22, 24, 25, 27, 28] Monte Carlo simulations incorporating all the scattering mechanisms suggest that grain boundary effects are critical in explaining the reduction in $\kappa$, particularly for thicker TG samples. These findings contribute to a deeper understanding of thickness-dependent phonon transport in graphite nanostructures and provide valuable insights for its application in thermal management of electronics.

## ASSOCIATED CONTENT

### Supporting Information

The Supporting Information is available free of charge on the ACS Publications website at DOI: XXX.

Thin graphite ribbon preparation; device fabrication and graphite ribbon transfer; details of thermal transport measurements of thin graphite ribbons; uniformity of graphite ribbon thickness; SEM images and surface profiles of thin graphite ribbons; analysis of heat loss through residual air and radiation; uncertainty analysis; heating power dependent temperature rise of heater/thermometer membrane with and without graphite ribbon; optical image of Pt ribbon device; first principles calculations; electron back scattering diffraction measurement; effect of point defect scattering.

## AUTHOR INFORMATION

**Corresponding Authors:**

Sangyeop Lee: sylee@pitt.edu

Renkun Chen: rkchen@ucsd.edu

**ORCID**

Renkun Chen: 0000-0001-7526-4981

Sangyeop Lee: 0000-0002-8783-9380

Lucas Lindsay: 0000-0001-9645-7993

Wonjae Jeon: 0000-0002-9325-6763

Yu Pei : 0009-0004-5753-7041

Xun Li: 0000-0001-8531-6140

**Author Contributions**

WJ, YP, XL contributed equally to this work. WJ and YP carried out the experiments, XL performed the simulations. RC, LL, SL supervised the project. WJ, YP, XL co-wrote the manuscript. All the authors contributed to the editing of the manuscript.

**Notes**

The authors declare no competing financial interest.

**ACKNOWLEDGMENTS**

Device fabrication and thermal transport measurement (R.C.) were support by National Science Foundation (NSF, DMR-2005181). Calculation guidance and manuscript development (L.L.) were supported by the U.S. Department of Energy, Office of Science, Office of Basic Energy Sciences, Material Sciences and Engineering Division. S.L. acknowledge the support from NSF (No. 1943807) and US Army (No. W911NF2320022). This work was performed in part at the San Diego Nanotechnology Infrastructure (SDNI) of UCSD, which was supported by NSF (ECCS-2025752).

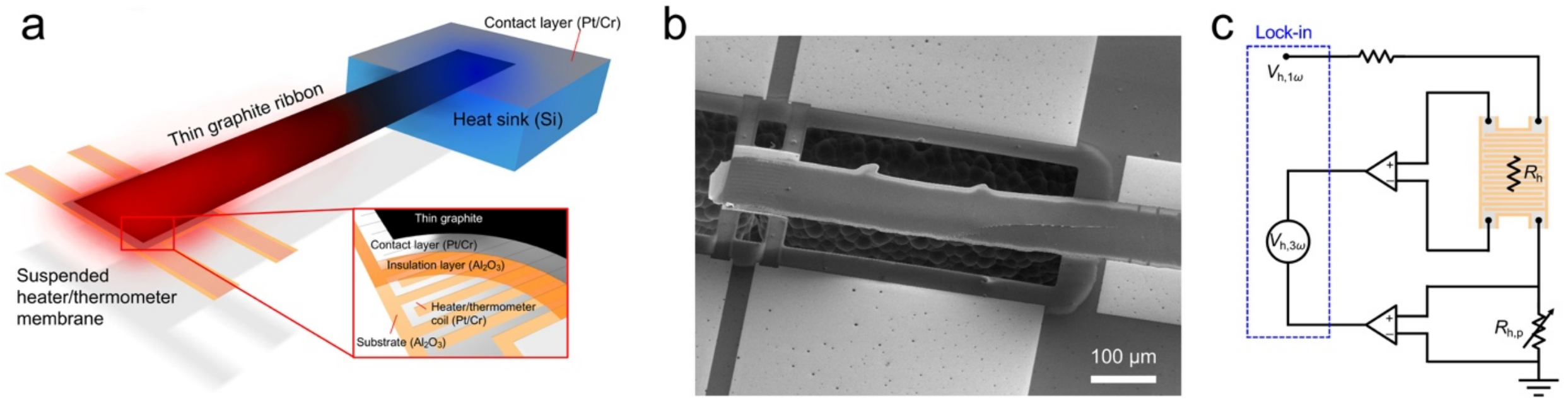

**Figure 1. Thermal conductivity measurement device for thin graphite ribbons.** (a) Schematic illustration and (b) SEM image of the device showing a suspended thin graphite ribbon (sample ID: TG-527 – see Table I) between the suspended heater/thermometer and the heat sink. The specific structure of the heater/thermometer is shown in the inset of figure (a). (c) Diagram of AC modulated heating of the four-probe $3\omega$ measurement system, where the heater is heated by a $1\omega$ current through the longest path of the heating coil, and the corresponding resistance change ($\Delta R_h$) due to the temperature rise is measured from the $3\omega$ voltage through the shortest path of the heating coil.

**Table 1. Sample dimensions of the measured thin graphite ribbons**

| Samples | Thickness (nm) | Width (μm) | Length (mm) | Peak Temperature (K) |
|---|---|---|---|---|
| TG-234 | 234 | 103 | 0.5 | 205.3 |
| TG-301 | 301 | 117 | 0.5 | 190.1 |
| TG-319 | 319 | 118 | 0.5 | 189 |
| TG-527 | 527 | 118 | 0.5 | 179.5 |

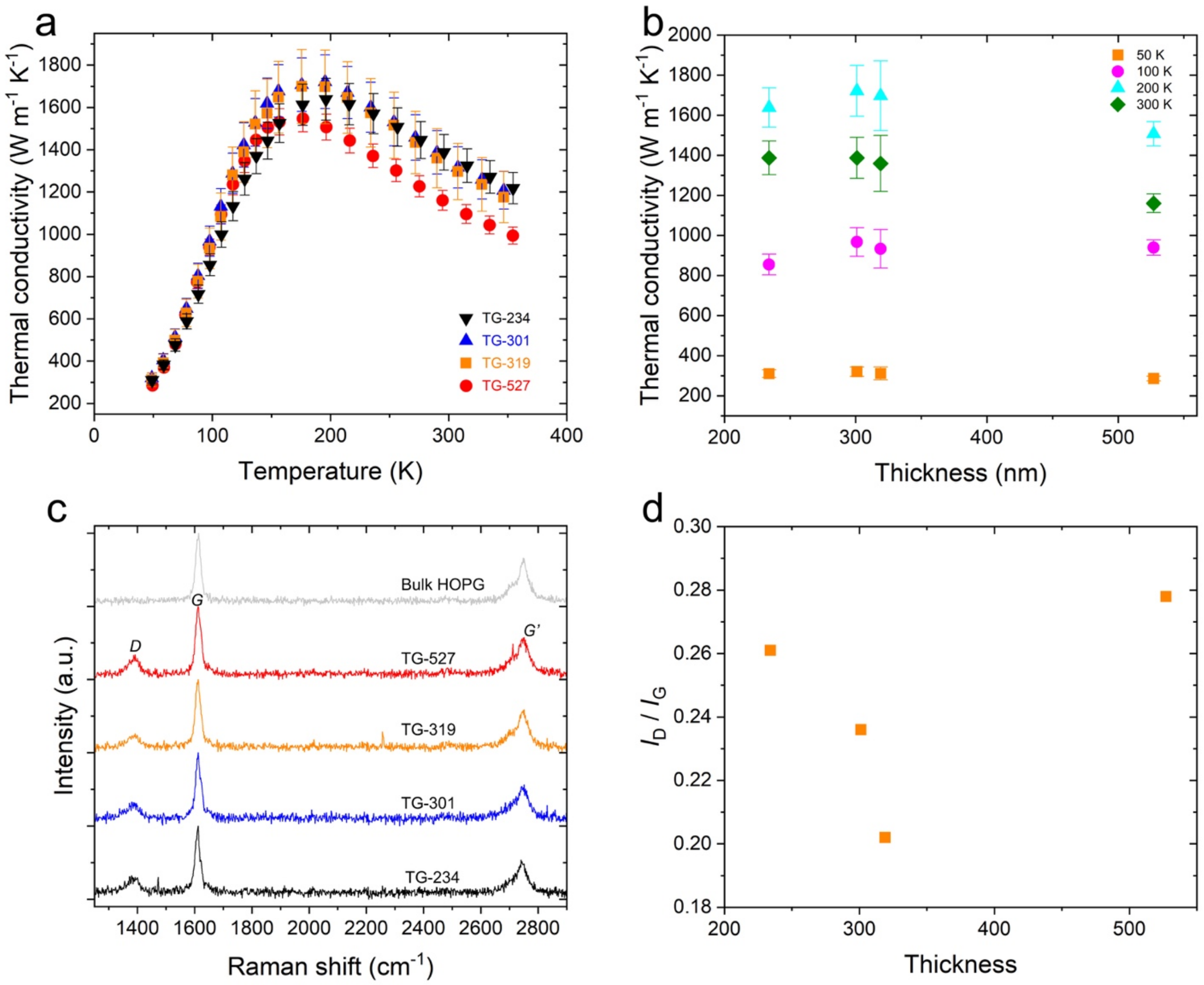

**Figure 2. Thermal transport properties of the measured thin graphite ribbons.** (a) Temperature dependence of thermal conductivity of the thin graphite ribbons. (b) Thickness dependence of thermal conductivity of the thin graphite ribbons at temperatures 50, 100, 200, and 300 K. Error bars for figures (a) and (b) indicate uncertainties from thermal conductance measurements and thin graphite ribbon dimension characterization. (c) Raman spectrum of TG 234-527 nm thick ribbons and bulk HOPG. (d) $I_D/I_G$ as a function of graphite thickness

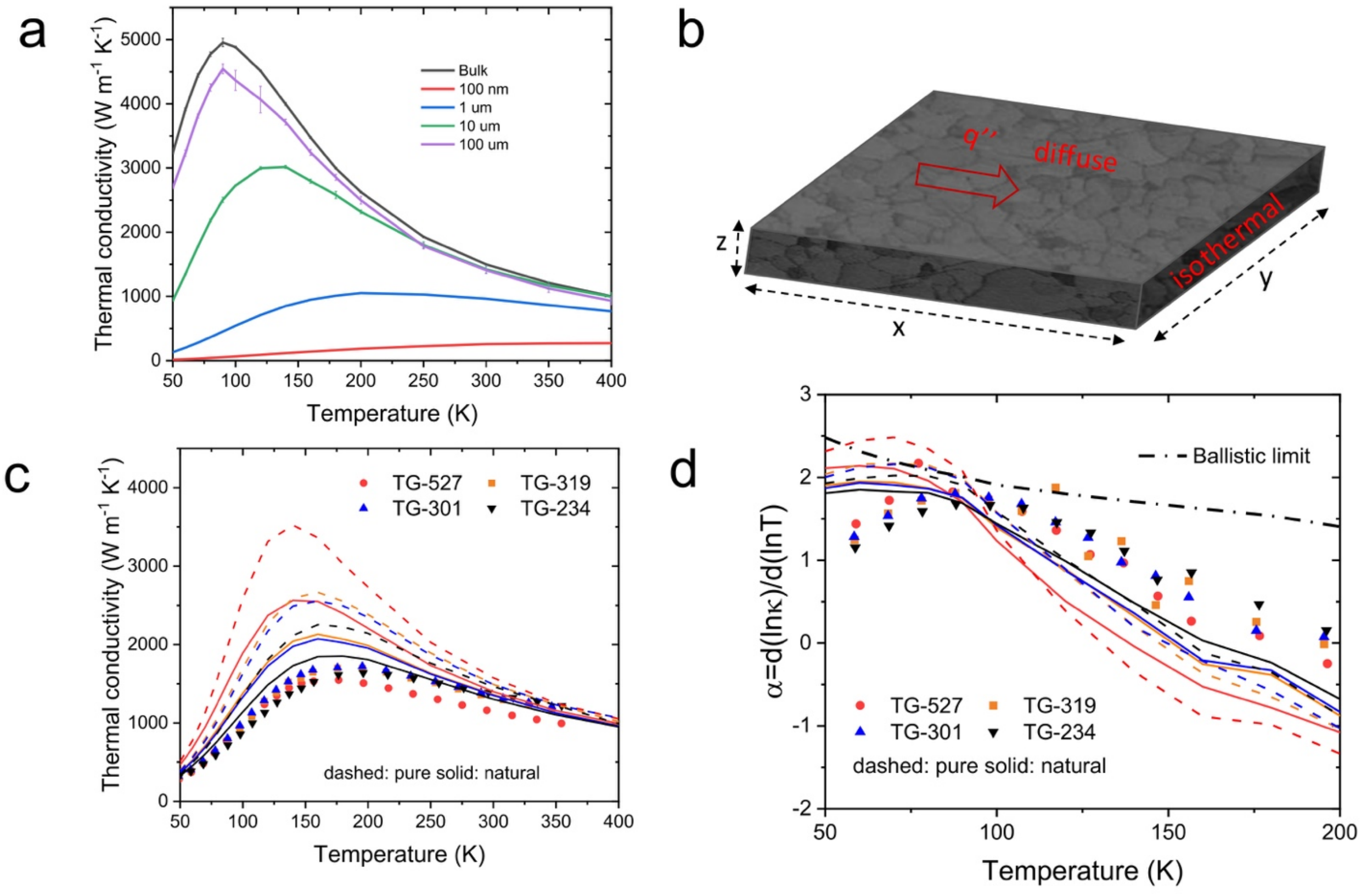

**Figure 3. Comparison between measurement and simulation for thin graphite samples** (a) Convergence test of thermal conductivity calculations on sample length for infinitely wide and thick graphite. (b) Schematic of heat flow and boundary conditions of three-dimensional graphite samples in the Monte Carlo simulation. (c) Calculated thermal conductivity for pure and natural samples. (d) Calculated temperature dependence $\alpha = d(ln\,\kappa)/d(ln\,T)$.

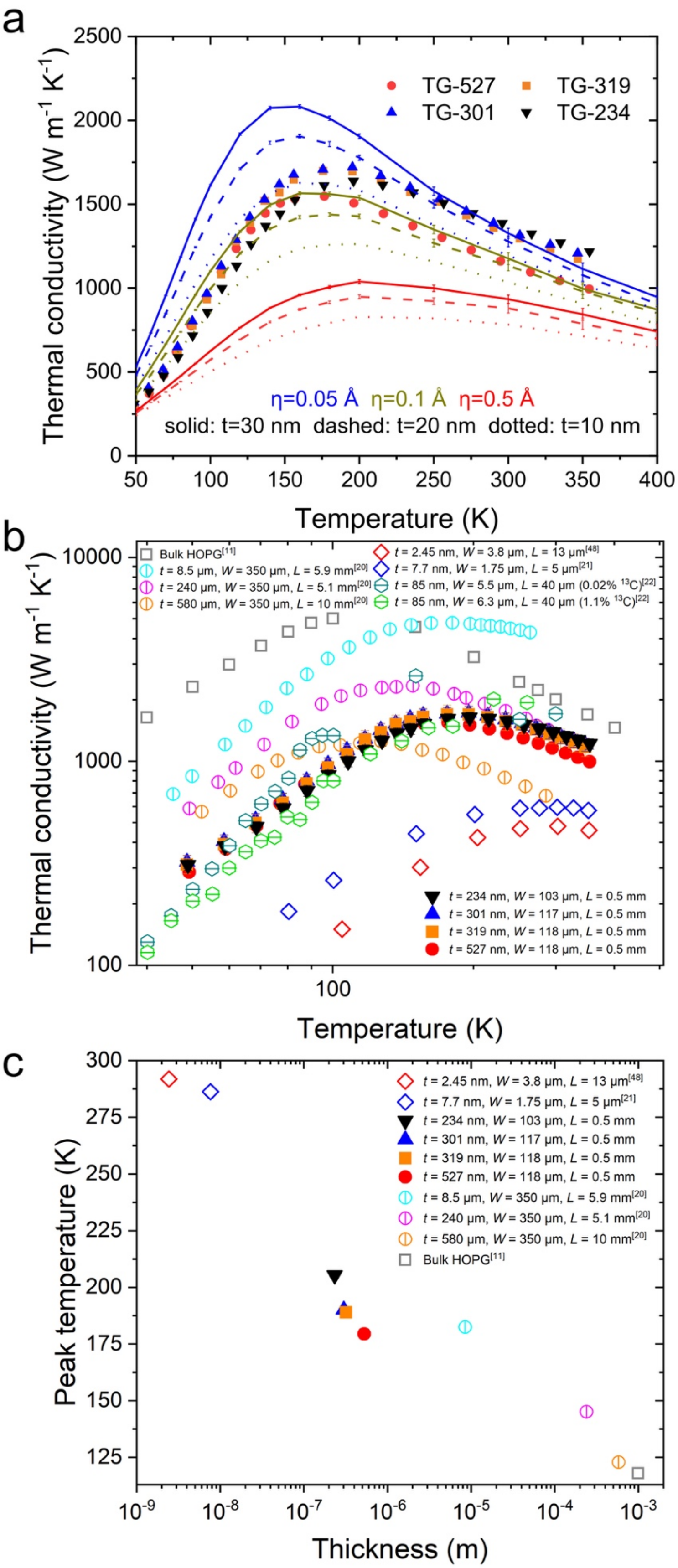

**Figure 4. Comparison of thermal transport properties of graphite with varying thickness** (a) Thermal conductivity as a function of temperature calculated with grain boundary scattering. (b) Comparison of the thermal conductivity results for the measured thin graphite ribbons (234 – 527 nm thickness) with reported measurements of various thicknesses of HOPG. The grey open square represents bulk HOPG, while the open diamonds represent HOPG with thicknesses of 2.45 nm (red) and 7.7 nm (blue), respectively. Thin graphite ribbons with thicknesses of 8.5 μm, 240 μm, and 580 μm (open circles with vertical lines) were measured using a thermocouple method. Open hexagons with horizontal lines represent a comparative study of isotopically purified and natural graphite with a thickness of 85 nm. (c) Thickness dependence of peak temperature of the measured graphite ribbons compared with various thicknesses of HOPG (2.45 nm, 7.7 nm, 8.5 μm, 240 μm, 580 μm), and 1 mm bulk HOPG. The symbols are the same as for figure (b).